# Do stable performance metrics guarantee stable model predictions?: An empirical investigation using the GUSTO-1 dataset

Natthanaphop Isaradech[1,2], Wachiranun Sirikul[1,3], Wuttipat Kiratipaisarl[1], Pakpoom Wongyikul[1], Noraworn Jirattikanwong[2], Phichayut Phinyo[2,*]

## Affiliations

[1]Department of Community Medicine, Faculty of Medicine, Chiang Mai University, Thailand, 50200

[2]Department of Biomedical Informatics and Clinical Epidemiology (BioCE), Faculty of Medicine, Chiang Mai University, Thailand, 50200

[3] Center of Data Analytics and Knowledge Synthesis for Health Care, Faculty of Medicine, Chiang Mai, Thailand 50200

## *Corresponding author

Phichayut Phinyo, MD., Ph.D.

Department of Biomedical Informatics and Clinical Epidemiology, Faculty of Medicine, Chiang Mai,

Thailand 50200, **Email:** phichayutphinyo@gmail.com; phichayut.phinyo@cmu.ac.th

**Abstract word counts:** 250

**Manuscript word counts:** 3362

**Number of figures:** 5

**Number of tables:** 2

**Reference numbers:** 46

## Abstract

**Background:** Model validation is standard practice to estimate and correct for model's overfitting in performance metrics. However, model validation leaves a gap regarding model stability when developed on slightly different training samples. While clinicians and model developers predominantly focus on stability in discrimination performance, relying solely on such metrics may mask significant prediction-level instability.

**Objective:** To investigate whether stable discrimination performance guarantee stability in individual model predictions, and to compare these patterns between logistic regression (LR) and artificial neural network (ANN).

**Methods:** We conducted an empirical comparison study using the GUSTO-I trial dataset (n = 40,830) with 30-day mortality as the outcome. Stratified random samples of six scenarios with different event per variables (EPV). For each scenario, LR and ANN were used to develop prediction models based on eight predictors. Two hundred bootstrap resamples were generated per scenario to estimate discrimination performance stability, mean absolute prediction error (MAPE), and classification instability index (CII).

**Results:** For LR, discrimination performance was stable when EPV was 44.50, while MAPE and CII did not stabilize until EPV reached 89.12. For ANN, discrimiantion performance was much more volatile. ANN's optimism stabilized at EPV = 178.25. ANN also produced severely miscalibrated predictions at EPV = 4.50.

**Conclusion:** Stable discriminative performance metrics may not guarantee stability in individual predictions. This dissociation was observed for both LR and ANN, although ANN models exhibited greater prediction instability. We recommend that model developers assess and report prediction-level stability, as discrimination performance may not adequately reflect the reliability of individual predictions.

**Keywords:** *Predictive Learning Models; Validation Studies; Discrimination Performance; Logistic Regression; Neural Network Models*

## Background

Clinical Prediction Models (CPM) are statistical or machine learning tools designed to estimate the probability of a specific health outcome or state that is present (diagnostic) or likely to happen in the future (prognostic) in individual patients based on their clinical characteristics and risk factors (1). Recent estimates indicate that nearly 250,000 articles reporting the development of CPMs across all medical fields were published up until 2024 (2). As a result, CPM evaluation is crucial to understand if the model could be trusted and to quantify the degree of uncertainty in its predictions before it is deployed in a real clinical setting (3,4). Usually, CPMs are evaluated by using performance metrics such as accuracy, discrimination and cilibration. By far, external validation is considered the gold standard for evaluating CPMs as it proves how well the model work in new unseen data, but the availability of an external dataset is usually limited by the resources of the researchers. Consequently, model developers typically decide to maximize the dataset's utility by testing the model performance by dividing the retrieved datset into train-test-splits (5). However, this may hide a CPM's overfitting nature because if the test set is small or captured during a specific timeframe that doesn't represent the whole population, the results might be overly optimistic or fluctuate due to random chance (6).

The current evidence suggested that internal validation should always be attempted for any proposed prediction model (7). This involes resampling methods such as bootstrapping or k-fold cross-validation to assess the difference between apparent model performance metrics and among the altered dataset variations (6,8–10). Among internal validation techniques, bootstrapping is widely preferred over other resampling owing to its superior data efficiency and lower bias (11). The primary goal is to quantify for model's "Optimism" which is the difference between the apparent performance and the performance assessed on the resampled dataset (12,13). When model performs worse on unseen data, high optimism, it is considered overfit (14,15). Subsequently, external validation is needed to assess its generalizability prior to clinical implementation (10,16).

However, model validation still leaves a gap in CPM evaluation. Riley and colleagues recently raised concern about the stability of CPMs when developed on a different slightly different training sample (17). Across models developed on the varied training samples, stability in discrimination performance was evaluated using the variability of area under the receiver operating characteristic curve (AUROC) or c-statistics, and individual prediction stability was assessed by the consistency of a patient's risk estimates (18). In practice, clinicians predominantly focus on overall discrimination performance, particularly the AUROC, though most clinical decisions are made for individual patients based on their specific predicted risk probabilities rather than group averages. Relying solely on the stability of discrimination performance metrics may overlook instability in the individual probabilities. Furthermore, the existing work on model stability has focused on logistic regression (LR). Complex algorithms' stability profile, such as artificial neural networks (ANNs), which have much larger parameter spaces and can converge to different configurations across resampled datasets, remains unclear (19–21).

To address this, we conducted an empirical analysis using the GUSTO-I trial dataset to investigate whether stable AuROC and low optimism from internal validation guarantee stable individual predictions. We developed models to predict 30-day mortality after admission using both standard regression and a complex machine learning approach across six sample size scenarios. We then compared discrimination performance stability and prediction instability metrics to assess the relationship between optimism stability and prediction-level stability.

## Methods

We conducted an empirical investigation using a real clinical dataset to preserve the complexity and multi-interactive nature of clinical data (22,23). Also, our study had included ANNs which are highly sensitive to noise and outcome distribution (24).

## Data source

We used the Global Utilization of Streptokinase and Tissue Plasminogen Activator for Occluded Coronary Arteries (GUSTO-I) trial dataset, which includes 40,830 patients with acute myocardial infarction (25). Eight candidate predictors including age, sex, history of hypertension, hypotension, and heart disease, ST-segment elevation, previous myocardial infarction, and systolic blood pressure were used to predict a binary event of 30-day mortality after admission of the patient.

### Sample size scenarios

To examine the effect of training sample size on discrimination performance stability and individual prediction stability, we created six subsets by stratified random sampling (preserving the outcome distribution) at fractions of 1.25%, 5%, 12.5%, 25%, 50%, and 100% of the full dataset, yielding sample sizes of 511, 2,042, 5,103, 10,208, 20,416, and 40,830, and event per variable (EPV) of 4.50, 17.88, 44.50, 89.12, 178.25, 356.38, respectively. These levels of sample size were selected to represent the sample size adequecy and overfitting risk in clinical prediction model, where EPV = 4.50 and 17.88 represents the underpowered scenario, EPV = 44.50 – 89.12 represents the adquately powered scenario and EPV over 178.25 represent high-powered scenarios (26,27).

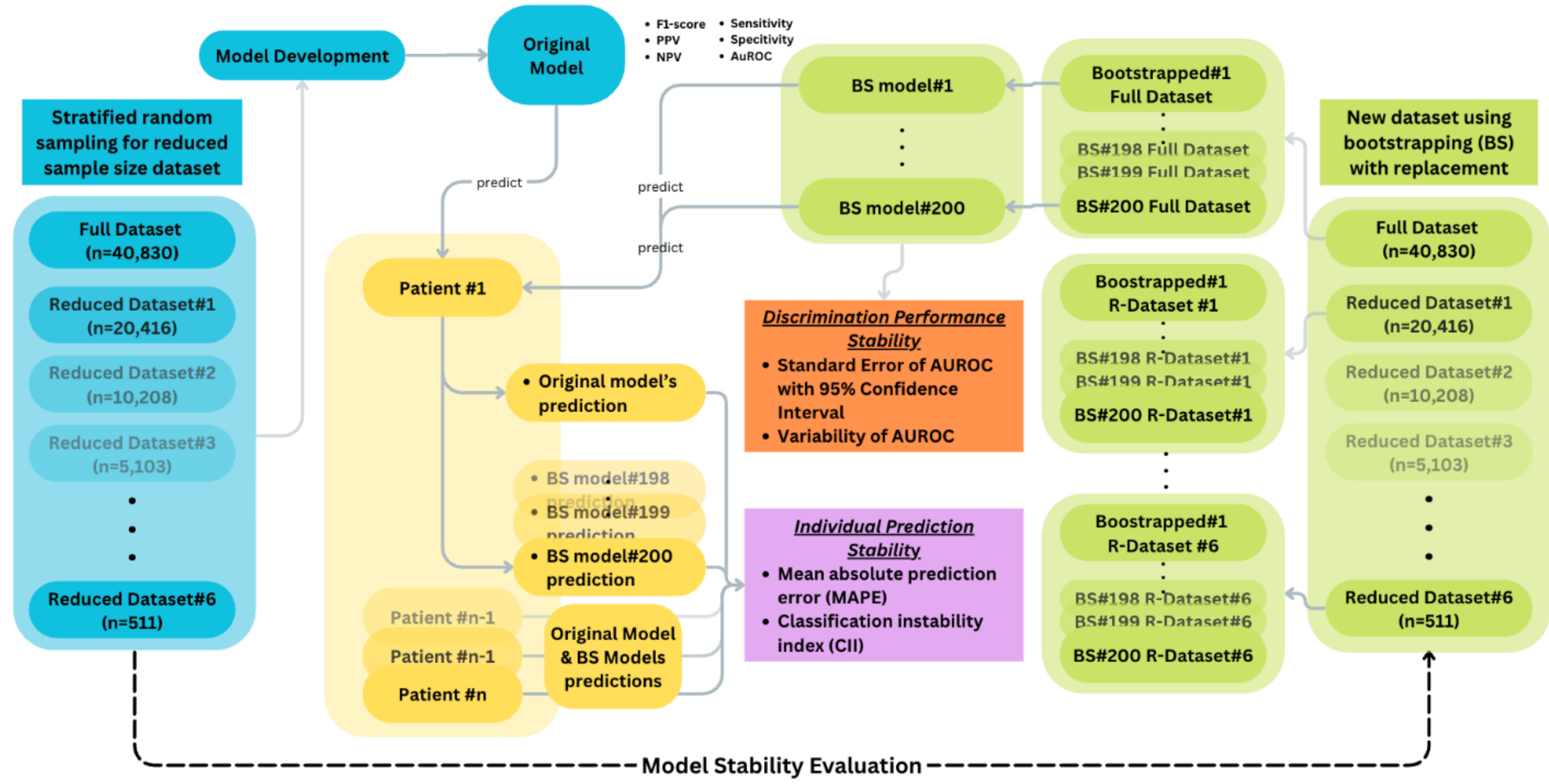

*Figure 1 Study Design Overview: Flowchart showing training sample stratification (fractions from 1.25% to 100% of GUSTO-I) and the bootstrap resample used to assess discrimination performance stability and individual prediction stability in logistic regression and ANN.*

## Model development

Two model types were developed for each sample size scenario.

**Logistic regression (LR)** was selected as it represents the established clinical gold standard. It is a parsimonious, transparent, and linear statistical classifier that is widely used and accepted in medical literature for a binary (dichotomous) outcome variable (28,29). The models were fitted with L2 regularisation, with the regularisation strength (C) and solver selected via 10-fold cross-validated grid search over C ∈ {0.001, 0.01, 0.1, 1, 10, 100, 1000} and solver ∈ {newton-cholesky, sag, saga, lbfgs} (30).

**Artificial neural network (ANN)** was selected as a representative of modern, flexible, and non-linear machine learning classifiers. ANNs are increasingly proposed for clinical risk prediction due to their capacity to capture complex high-dimensional relationships (31,32). ANNs could also be applied to develop text or image classification models (33,34). However, we assumed their flexibility and high parameter density made them potentially susceptible to overfitting and high instability to training data variability. The ANNs were implemented as multi-layer perceptrons classifier, with hyperparameters selected via 10-fold cross-validated grid search over multiple hidden layer architectures ∈ {(8,), (16,), (32,), (64,), (32, 16), (64, 32), (64, 32, 16)}, activation functions ∈ {logistic, tanh, relu}, learning rate schedules ∈ {constant, adaptive}, and regularisation strength (alpha) ∈ {0.0001, 0.001, 0.01, 0.1, 1, 10, 100, 1000}, with early stopping enabled and a maximum of 500 iterations.

**Assessment of stability in discrimination performance and stability in individual predictions**

To evaluate the performance and stability of the clinical prediction models, we implemented a bootstrap resampling approach to assess both discrimination performance and individual prediction stability as shown in **Error! Reference source not found.**. For each sample scenario, 200 bootstrap samples was used to train separate models using the same model training configuration (18). This yield 200 bootstrapped models. Each of which was then applied to the original dataset to generate predicted probabilities. The resulting 200 sets of predictions, one per bootsrapped model, were compared aginast the predictions from the original model model to derive discrimination stability metrics, the variability of AUROC across bootstrapped models, and inidividual prediction instability; mean absolute prediction error (MAPE) and classification instability index (CII) (17).

**Evaluation: Stability in discrimination performance and stability in individual prediction**

- Step 1: Train the original model on the original sample, and generate original estimated probabilities ($P_{original}$) for all patients in the original sample.
- Step 2: Draw 200 bootstrap samples with replacement from the original dataset.
- Step 3: For each of the 200 bootstrap samples:
    - Train a new model ($Model_b$) on the bootstrap sample.
    - Apply the fitted Model_b back to the original sample to generate estimated probabilities ($P_b$) for all patients.
    - Compute its AUROC (**Bootstrapped $AUROC_b$**) on the original sample, yielding a distribution of 200 bootstrapped AUROCs (**$AUROC_1$,$AUROC_2$,…,$AUROC_{200}$**).
    - For each $patient_i$ in the original sample:
        - Compute MAPE from the absolute difference between predicted probability ($P_{b,i}$) from the bootstrapped model and predicted probability from the original model ($P_{original,i}$)
        - Compute CII by checking if the classification status changes across the decision threshold (set at 10% or 0.10). A status change occurs if the patient is classified as high-risk by one model (probability >= 10%) and low-risk by the other (probability < 10%).
- Step 4: Compute the stability of discrimination performance
    - Compute Standard Error (SE) and 95% Confidence Interval of AUROC across the 200 AUROC values from models trained from bootstrapped samples.
    - Compute AUROC variability from the absolute difference between 200 bootstrapped AUROCs ($AUROC_1$,$AUROC_2$,…,$AUROC_{200}$) and their median value.

- Step 5: Summarize MAPEs and CIIs across all 200 bootstrap samples by calculating their mean values and their 95% confidence intervals (calculated using the 2.5$^{th}$ and 97.5$^{th}$ percentiles over the 200 values).

### Statistical Analysis

All metrics are reported as mean values of the metrics (AUROC, MAPE, and CII) alongside their 95% bootstrapped percentile confidence interval (CI), defined as the 2.5$^{th}$ and 97.5$^{th}$ percentiles of the distributions across the 200 bootstrap iterations (35). Analyses were conducted in Python 3.14.3 using scikit-learn (36), statsmodels (37), SciPy (38), and matplotlib packages (39).

## Results

The baseline characteristics of the data derived from the GUSTO-I cohort (N = 40,830) were analyzed. The primary outcome, 30-day mortality, occurred in 2,851 patients, corresponding to a baseline event rate of 6.98%. The study cohort had a mean age of 60.91 years (SD = 11.94, range: 19–110 years) and was predominantly male (n = 30,552; 74.8%) compared to female (n = 10,278; 25.2%). In terms of clinical presentations and cardiovascular risk factors: Hypertension was documented in 25,310 patients (62.0%). Hypotension was documented in 3,378 patients (8.3%). Tachycardia was present in 13,394 patients (32.8%). Prior myocardial infarction was reported in 6,726 patients (16.5%). The cohort presented with a mean systolic blood pressure of 128.98 mmHg (SD = 23.89) and a mean number of leads with ST-segment elevation of 4.10 (SD = 1.90).

### Discrimination performance, and model stability

For LR, The mean bootstrap AUROC was 0.569 (95% CI: 0.479 - 0.619) at EPV = 4.50 (n = 511) and increased to 0.802 (95% CI: 0.812–0.816) at EPV = 44.50 (n = 5,103), with the full dataset (n = 40,830) yielding a mean AUROC of 0.720 (95% CI: 0.713 - 0.725). SE of AUROC across boostrapped samples decreased steadily with increasing sample size from 0.040 at n = 511 to 0.003 at n = 40,830. The range of 95% CI of AUROC stabilised at relatively small sample sizes and fair EPV; by EPV = 44.50 (n = 2,042) as shown in Table 1.

*Table 1 Logistic Regression: Discrimination, and Stability Metrics Across Sample Sizes*

| | AuROC | | | | MAPE | | CII | |
|---|---|---|---|---|---|---|---|---|
| **N (EPV)** | **original** | **mean** | **SE** | 95% CI | **mean** | **95% CI** | **mean** | **95% CI** |
| 511 (4.50) | 0.622 | 0.569 | 0.040 | 0.479 - 0.619 | 0.013 | 0.004–0.042 | 0.067 | 0.000–0.440 |
| 2,042 (17.88) | 0.802 | 0.785 | 0.012 | 0.758 - 0.802 | 0.015 | 0.001–0.065 | 0.052 | 0.000–0.044 |
| 5,103 (44.50) | 0.812 | 0.802 | 0.007 | 0.786 - 0.812 | 0.008 | 0.001–0.034 | 0.027 | 0.000–0.350 |
| 10,208 (89.12) | 0.798 | 0.791 | 0.005 | 0.779 - 0.798 | 0.006 | 0.001–0.022 | 0.022 | 0.000–0.265 |
| 20,416 (178.25) | 0.729 | 0.722 | 0.004 | 0.711 - 0.729 | 0.003 | 0.001–0.012 | 0.016 | 0.000–0.265 |
| 40,830 (356.38) | 0.725 | 0.720 | 0.003 | 0.713 - 0.725 | 0.002 | 0.000–0.006 | 0.008 | 0.000–0.100 |

**AUROC**; Area Under the Receiver Operating Characteristic curve, **CI**; Confidence Interval, **CII**; Classification Instability Index, **EPV**; Event per variable, **MAPE**; Mean Absolute Prediction Error, **SE**; Standard Error.

Mean MAPE was highest at EPV = 356.38 (n = 40,830) and lowest at EPV 17.88 (n = 2,042). Mean CII followed a similar declining trend, from 0.067 (95% CI: 0.000–0.440) at the lowest EPV ratio case to 0.008 (95% CI: 0.000–0.100) at the full dataset. **Error! Reference source not found.** illustruated that MAPE and CII remained substantially elevated and variable at underpowered sample sizes both metrics decreased progressively with increasing sample size, behaving comparablly to AUROC variability, even though it does not improve much after EPV = 44.50 (n = 5,103).

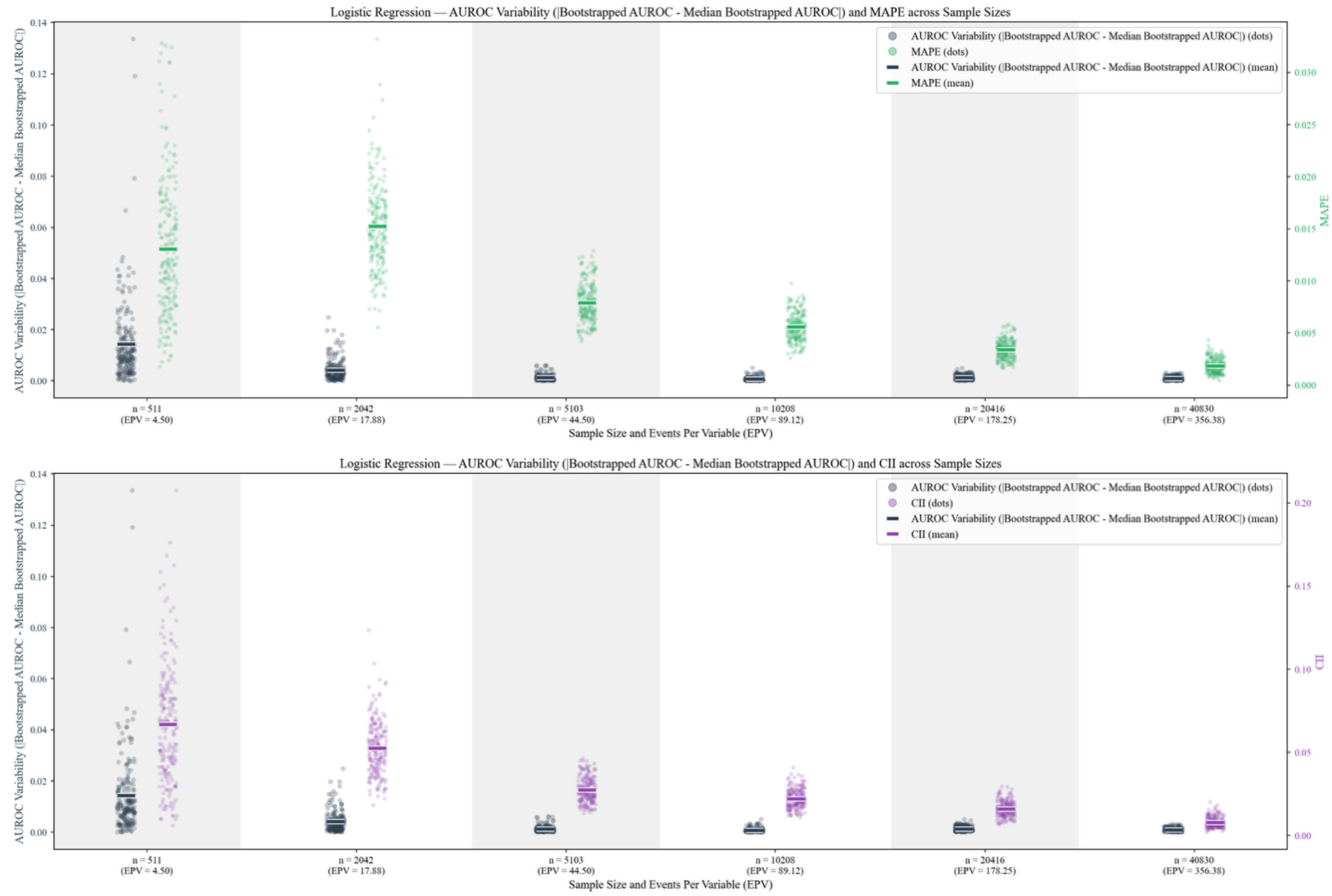

*Figure 2 Logistic Regression stability metrics across sample sizes. Relationship between AUROC variability (AUROC difference from median, left y-axis) and patient-level prediction instability (MAPE, top panel; CII, bottom panel, right y-axis) across six sample sizes.*

ANNs showed a different pattern in

Table 2. Mean AUROC values were generally lower and more variable, ranging from 0.599 (95% CI: 0.525 - 0.709) at n = 5,103 to 0.646 (95% CI: 0.318 - 0.695) at n = 10,208. Mean AUROC reached 0.772 (95% CI: 0.715 - 0.792) when sample size reaches 20,416. However, SE of AUROC for ANN did not decrease as sample size increases linearly. ANN's SE of AUROC was highest at EPV = 44.50 (n = 5,103).

*Table 2 Artificial Neural Network: Discrimination, and Stability Metrics Across Sample Sizes*

| | AUROC | | | | MAPE | | CII | |
|---|---|---|---|---|---|---|---|---|
| **N (EPV)** | **original** | **mean** | **SE** | **95% CI** | **mean** | **95% CI** | **mean** | **95% CI** |
| 511 (4.50) | 0.639 | 0.599 | 0.029 | 0.525 - 0.636 | 0.003 | 0.002–0.006 | 0.000 | 0.000–0.000 |
| 2,042 (17.88) | 0.570 | 0.518 | 0.042 | 0.429 - 0.568 | 0.030 | 0.003–0.101 | 0.074 | 0.000–0.325 |
| 5,103 (44.50) | 0.711 | 0.589 | 0.163 | 0.252 - 0.709 | 0.032 | 0.008–0.106 | 0.117 | 0.000–0.325 |
| 10,208 (89.12) | 0.696 | 0.646 | 0.082 | 0.318 - 0.695 | 0.025 | 0.006–0.060 | 0.120 | 0.000–0.685 |
| 20,416 (178.25) | 0.793 | 0.772 | 0.021 | 0.715 - 0.792 | 0.021 | 0.05–0.060 | 0.094 | 0.000–0.695 |
| 40,830 (356.38) | 0.732 | 0.725 | 0.005 | 0.714 - 0.732 | 0.019 | 0.005–0.067 | 0.104 | 0.000–0.770 |

**AUROC**; Area Under the Receiver Operating Characteristic curve, **CI**; Confidence Interval, CII; Classification Instability Index, **EPV**; Event per variable, **MAPE**; Mean Absolute Prediction Error.

At EPV lower than 178.25 (n = 20,416), ANN showed notable instability with MAPE ranging from 0.003 to 0.032 and CII ranging from 0.000 to 0.120. The values are higher than those observed for logistic regression at equivalent sample sizes. Critically, even at the full dataset (EPV = 356.38, n = 40,830), ANN models showed a mean MAPE of 0.019 (95% CI: 0.005–0.067) and CII of 0.104 (95% CI: 0.000–0.770). **Error! Reference source not found.** illustrates how AUROC variability, improved as sample size increases for ANN except for MAPE and CII but unlike LR, they did not decrease linearly. Even at full dataset, ANN prediction prediction instability metrics are still highly variable when compared to LR.

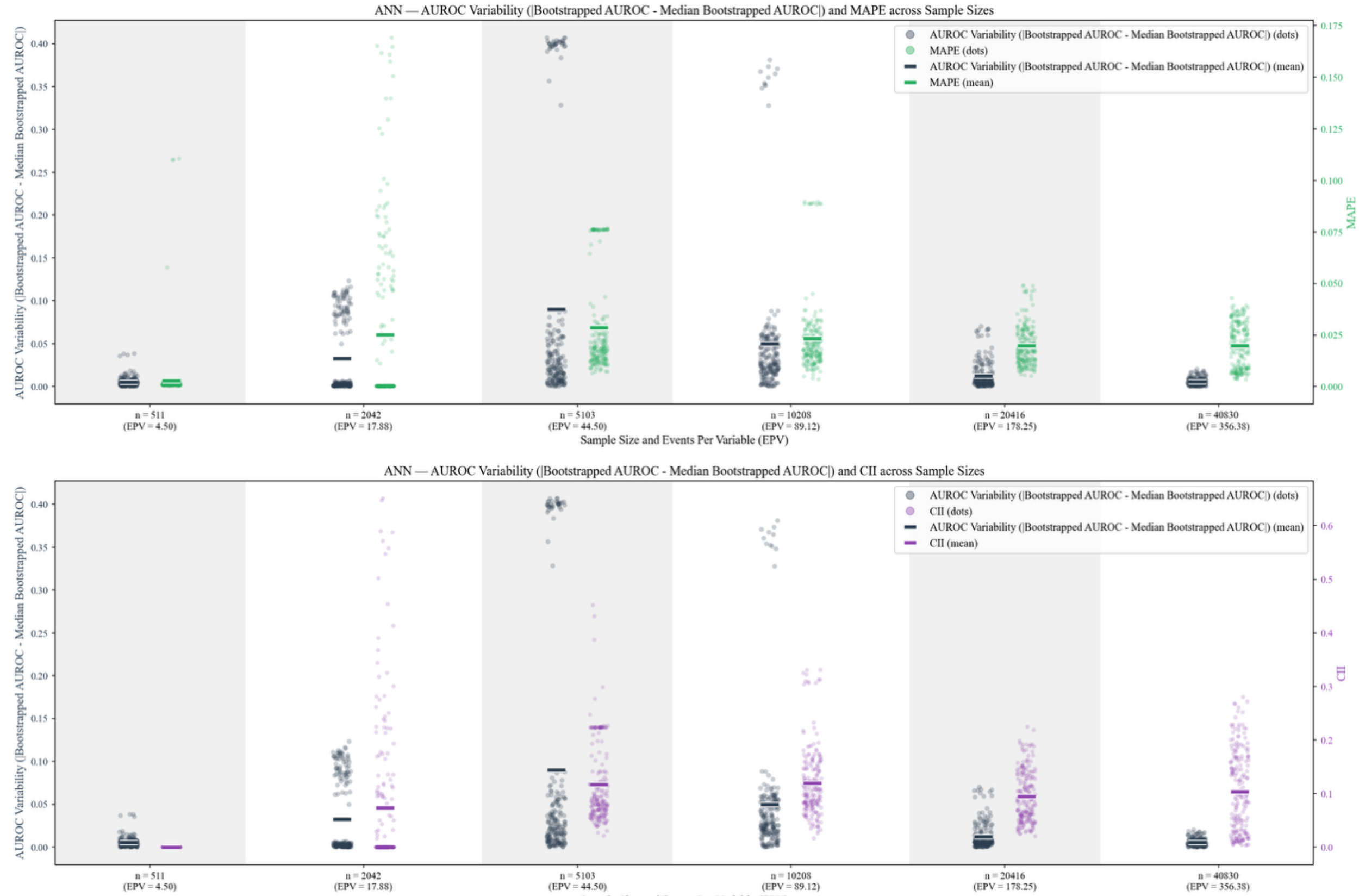

*Figure 3 ANN stability metrics across sample sizes. Relationship between AUROC variability (AUROC difference from median, left y-axis) and patient-level prediction instability (MAPE, top panel; CII, bottom panel, right y-axis) across six sample sizes.*

## Prediction Instability in Complex Models

We explored the difference of prediction model stability on a more complex model with different optimization techniques, ANN, in addition to LR. The results showed that the magnitude of ANN prediction instability are largely higher than from LR model. Even at full dataset, ANN prediction prediction metrics are still highly unstable.

ANN's prediction stability was substantially more volatile and did not follow a gradually declining trend as sample size increases. At EPV = 4.50 (n = 511), the ANN model produced severely miscalibrated predicted probabilities that were compressed into a narrow range and sat entirely above the 10% decision threshold (MAPE = 0.003, CII = 0.000) as illustrated in **Error! Reference source not found.**.

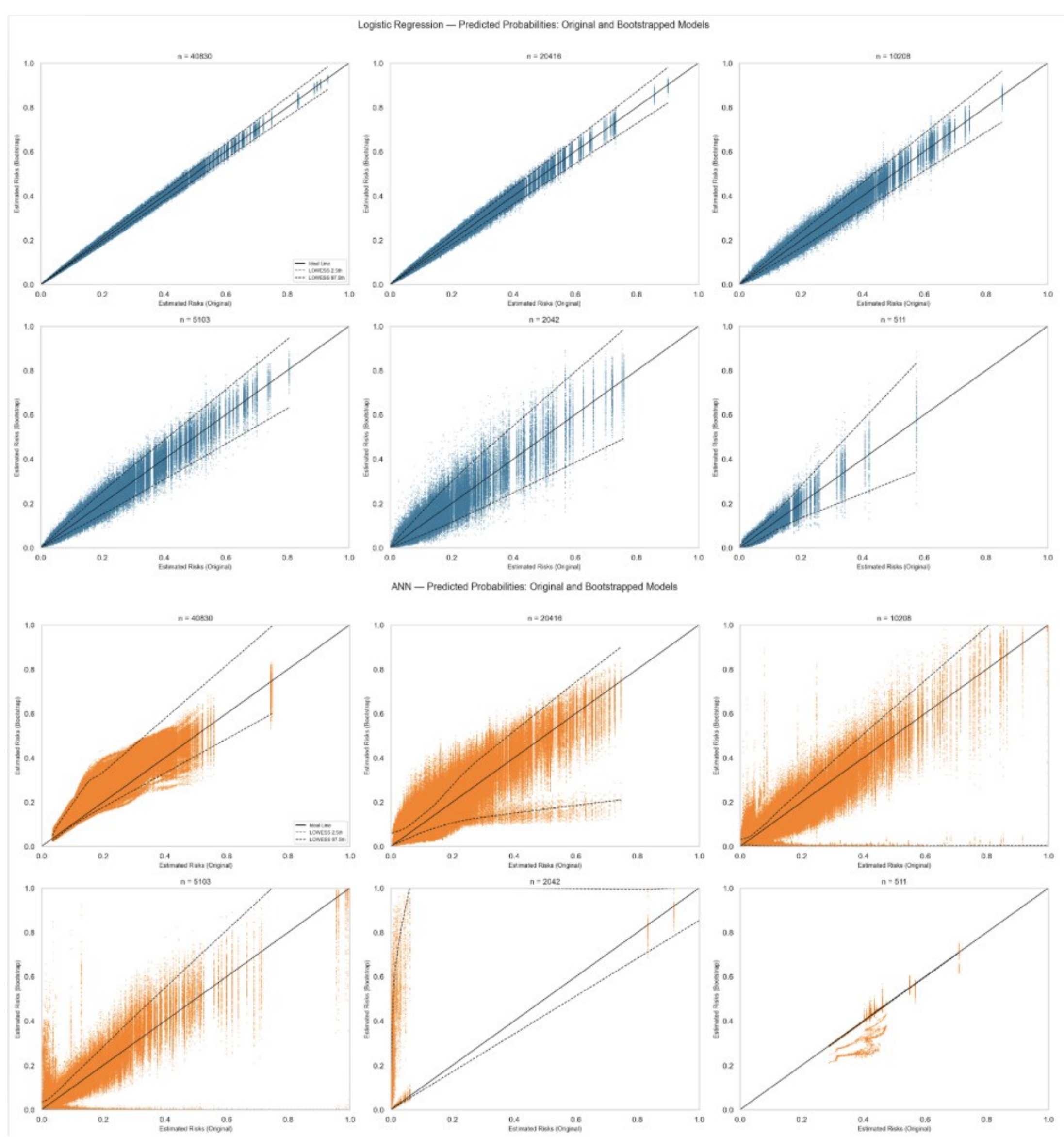

*Figure 4 Predicted risk probabilities between original and bootstrapped models. Original predicted risk (x-axis) vs. predicted probabilities from 200 bootstrapped models (y-axis) across six sample sizes for Logistic Regression (blue, top) and ANN (orange, bottom)*

## Discussion

This empirical study demonstrated that stable discrimination performance might not guarantee stable individual model predictions. Discrimination performance stability, measured by the SE of the AUROC, its 95% confidence intervals and variability of AUROC, evaluates the volatility of overall model discriminative performance across resampled datasets, while prediction stability metrics, such as MAPE and CII, measure the variance in absolute risk probabilities for individual patients. This suggested that these metrics are distinct dimensions of a CPM and represent different levels of CPM stability, where discrimination stability is considered the primary level of CPM stability, as the results showed that overall discrimination performance stabilized at lower sample sizes (or lower EPV ratios) before individual prediction stability in both models. In binary LR, both the SE of the AUROC and prediction instability metrics decreased and stabilized as the sample sizes (and EPV ratios) increased, showing a highly coordinated improvement(27,40). However, that might not be the case for complex models such as ANN (41,42).

We assumed the following reasons why a complex model like ANN dealed with prediction stability poorly; First, ANNs have a substantially larger parameter space than logistic regression (43). Even the simplest architecture in our grid search, a single hidden layer with 8 units, involves more parameters than a logistic regression model with 8 predictors. More complex architectures (e.g., 64–32–16) involve hundreds of parameters. With more parameters to estimate from the same data, each bootstrap resample produces a different solution, leading to higher variability in individual predicted probabilities. Second, the ANN loss surface is non-convex, containing multiple local minima points. Unlike LR, which has a unique global optimum for a given dataset, ANN training with stochastic gradient descent can converge to different local minima depending on the training data composition, random initialisation, and learning rate. When the training sample changes the optimiser may end up in different regions of the parameter space, causing different prediction functions even though the aggregate AUROC looks similar (44). Lastly, because of the size of the parameters, optimization techniques, and the complications of ANNs, we assume that they would need a significantly bigger sample size to reach both optimism and prediction-level stability (26,45).

In clinical practice, the adoption of CPMs is heavily driven by discrimination performance, specifically the AUROC. Clinicians frequently rely on the apparent AUROC presented in the studies, and often overlook the fundamental difference between apparent performance and internally or externally validated performance. Moreover, when clinicians do address a model's reliability, they typically focus on discrimination performance stability, such as the standard error or confidence intervals of the AUROC. A model with stable discrimination performance is usually considered validated and clinically robust.

However, our study showed a case where even when discrimination performance was stable, it did not mean the model can be trusted. When a model's individual predictions are unstable, the consequences are clinically severe. A patient's estimated risk of 30-day mortality might be 75% in one bootstrap iteration but 55% in another. The decision to treat or not treat becomes dependent on the random variation of the cohort used to train the model, rather than the patient's actual clinical state (18). If two clinicians run slightly different versions or updates of the same model on the same patient and get different treatment recommendations, the clinical value of model is compromised, regardless of how high the model's reported AUROC is as illustrated in **Error! Reference source not found.**.

*Figure 5 Prediction instability and its clinical consequence: Same patient prediction (true risk = 60%), with five bootstrap (BS) model replicates using the same training strategy and hyperparameters, and cutoff = 0.50*

Several limitations should be acknowledged. First, this study used a single dataset (GUSTO-I) with a single binary outcome (30-day mortality), and the generalisability of our findings to other clinical contexts, outcome types, or datasets with different predictor structures warrants further investigation. Though, we believe that GUSTO-I's large sample size provided a high-quality benchmark to represent subsampling across a wide range of EPV scenarios. Second, we examined only two specific modelling approaches; the behaviour of other algorithms such as tree-based classifiers may differ. However, these two models were intentionally selected to represent the widely-used approaches: traditional statistical models (LR) and complex, non-linear neural networks (ANN). Third, the GUSTO-I dataset has a relatively low event rate, which may amplify instability at small sample sizes; datasets with more balanced outcomes might show different patterns. Future work should examine how class-rebalancing techniques may affect model prediction stability. Fourth, our machine learning implementation used a specific hyperparameter search optimization, and different architectures or training strategies. Future studies could examine whether different model configurations would result in different stability profiles.

## Conclusion

Stable discrimination performance metrics may not ensure stability in individual model predictions. This study demonstrated a dissociation between discrimination performance stability and individual prediction-level stability in both LR and ANN. These findings suggest that discrimination performance stability and individual prediction-level stability capture distinct aspects of model reliability. We recommend that model developers evaluate and report prediction-level stability metrics, such as MAPE and CII, alongside discrimation performance metrics, particularly when working with limited sample sizes or complex algorithms.

## Declaration

**Author Affiliations:** [1]Department of Community Medicine, Faculty of Medicine, Chiang Mai University, Thailand, [2]Department of Biomedical Informatics and Clinical Epidemiology (BioCE), Faculty of Medicine, Chiang Mai University, Thailand

**Competing interests:** None.

**Contributions:** **N.I.** and **P.P.** initiated the project. **W.K., P.W.**, and **N.J.** participated in the study design and manuscript writing. **N.I.** carried out the modeling work, wrote the original draft of the manuscript, and conducted manuscript editing. **W.S.** contributed to the modeling framework. **W.S.** and **P.P.** provided overall supervision and reviewed/edited the manuscript. All authors have read and agreed to the published version of the manuscript.

**Acknowledgments:** The authors thank the GUSTO-I trial investigators and participants for making the dataset publicly available for research. We also thank our respective departments and faculties for providing the research facilities and resources that supported this study.

**Funding**: This research received no external funding.

**Disclosure of Delegation to Generative AI:** The authors declare the use of generative AI in the research and writing process. According to the GAIDeT taxonomy (2025), the following tasks were delegated to GAI tools under full human supervision (46): Code generation, Code optimization, Process automation, Creation of algorithms for data analysis, Proofreading and editing, Translation. The GAI tool used was: Gemini Flash 3.5, and Claude Opus 4.6. Responsibility for the final manuscript lies entirely with the authors. GAI tools are not listed as authors and do not bear responsibility for the final outcomes. Declaration submitted by: Natthanaphop Isaradech.

**Availability of data and materials**: The code supporting the findings of this study is available in the GitHub repository: https://github.com/natthanaphop-isa/model_instability.git.